\documentclass[aps,pre,twocolumn,superscriptaddress,10pt]{revtex4-2}

\usepackage[utf8]{inputenc}
\usepackage{amsfonts}
\usepackage{amsmath}
\usepackage{amssymb}
\usepackage{graphicx}
\usepackage[colorlinks=true, allcolors=blue]{hyperref}

\usepackage{multirow}

\newcommand{\NNseq}{NN$_\text{seq}$ }
\newcommand{\NNpar}{NN$_\text{par}$ }

\newcommand{\NNs}[2]{\text{NN}$_{#1}^{#2}$}
\newcommand{\NNp}[1]{\text{NN}$_{\Delta\mu}^{#1}$}

\begin{document}

\title{Neural surrogates for crystal growth dynamics with variable supersaturation: explicit vs. implicit conditioning}

\author{Matteo Rigoni}
\affiliation{Dept. of Materials Science, University of Milano-Bicocca, 20125 Milano, Italy}

\author{Daniele Lanzoni}
\affiliation{Dept. of Materials Science, University of Milano-Bicocca, 20125 Milano, Italy}
\affiliation{Dept. of Physics, University of Genova, 16146 Genova, Italy}

\author{Francesco Montalenti}
\affiliation{Dept. of Materials Science, University of Milano-Bicocca, 20125 Milano, Italy}

\author{Roberto Bergamaschini}
\affiliation{Dept. of Materials Science, University of Milano-Bicocca, 20125 Milano, Italy}
\email{roberto.bergamaschini@unimib.it}\makeatletter
\def\maketitle{
\@author@finish
\title@column\titleblock@produce
\suppressfloats[t]}
\makeatother

\begin{abstract}
\noindent Simulations of crystal growth are performed by using Convolutional Recurrent Neural Network surrogate models, trained on a dataset of time sequences computed by numerical integration of Allen-Cahn dynamics including faceting via kinetic anisotropy. Two network architectures are developed to take into account the effects of a variable supersaturation value. The first infers it implicitly by processing an input mini-sequence of a few evolution frames and then returns a consistent continuation of the evolution. The second takes the supersaturation parameter as an explicit input along with a single initial frame and predicts the entire sequence. The two models are systematically tested to establish strengths and weaknesses, comparing the prediction performance for models trained on datasets of different size and, in the first architecture, different lengths of input mini-sequence. The analysis of point-wise and mean absolute errors shows how the explicit parameter conditioning guarantees the best results, reproducing with high-fidelity the ground-truth profiles. Comparable results are achievable by the mini-sequence approach only when using larger training datasets. The trained models show strong conditioning by the supersaturation parameter, consistently reproducing its overall impact on growth rates as well as its local effect on the faceted morphology. Moreover, they are perfectly scalable even on 256 times larger domains and can be successfully extended to more than 10 times longer sequences with limited error accumulation. The analysis highlights the potential and limits of these approaches in view of their general exploitation for crystal growth simulations.
\end{abstract}


\maketitle

\section{Introduction}\label{sec:introduction}

\noindent Machine Learning (ML) and particularly Deep Learning (DL) is currently revolutionizing materials science, both from an experimental and a computational standpoint~\cite{wei2019machine, choudhary2022recent, schmidt2019recent, peivaste2025artificial}. In the latter context, one of the perhaps most intriguing possibilities brought by data-driven approaches is the significant reduction of the computational costs associated with simulations, with the possibility of screening new materials and processes faster than ever before~\cite{lyngby_data-driven_2022, zhao_physics_2023, chenebuah_deep_2024, karpovich_deep_2024}, with potential disruptive impacts on the speed of technological and fundamental discoveries.

One of the applications that could most benefit from the use of ML methods are continuum models for the temporal evolution of materials. This class of simulations allows the study of nano- to macro-scale processes, such as crystalline materials growth, microstructure formation and evolution, mechanical and electrical responses, etc. Among these applications, problems involving interface motion are particularly challenging, since they often require computationally expensive numerical procedures and the solution of challenging coupled PDEs. One of the traditional ways to tackle this kind of issue is through phase-field (PF) methods~\cite{provatas_phasefield_2010, liCCP2009}, in which the description of the system geometry or microstructure is traced implicitly by the evolution of one or multiple order parameters. Still, PF approaches could be nonetheless expensive, due to the requirement of fine meshes and of the solution of stiff nonlinear equations. Indeed, several ML methods have been proposed in recent years to mitigate these issues. These have exploited many of the state-of-the-art methods in DL, such as autoencoders for the so-called "latent space models"~\cite{montesNPJCM2021, huCMAME2022, oommenNPJCM2022, tepACTAMATER2025}, Convolutional Neural Networks (CNNs)~\cite{alhadaNPJCM2024, bonneville2026towards}, neural operators~\cite{oommenNPJCM2022, peivaste2025teaching}, and graph NN~\cite{fanMLST2024} to cite a few. A comprehensive review can be found at Ref.~\cite{lanzoniJPCM2025}. Among other approaches, the use of Convolutional Recurrent Neural Networks (CRNN)~\cite{shiARXIV2015, BallasICLR2016} is particularly interesting, as they allow to jointly capture spatial (Convolutional aspect) and temporal (Recurrent aspect) correlations in data. Several studies have already shown how CRNN may be used to learn and reproduce at a smaller computational cost simulations of the evolution of technologically and theoretically relevant mesoscale phenomena \cite{yangPATTERNS2021,wuCMS2023, lanzoniMLST2024, fantasiaADVTHESIM2026}, such as grain growth, solidification processes, and spinodal decomposition. Moreover, the fully-convolutional nature of the approach allows for generalizations to arbitrary domain size, increasing the range of applicability of these models.

One of the peculiarities of Recurrent NNs is their possibility of both generating and processing temporal sequences. The second property has been explicitly used in some works exploiting CRNN to predict the microstructural and morphological evolution of materials based on short initial sequences~\cite{farizhandiSCIREP2022, wuCMS2023}. Perhaps more importantly, this possibility would be particularly interesting in experimental contexts and situations in which the dynamics is driven by hidden or unknown factors. Moreover, instead of explicitly providing the CRNN model with all the quantity determining the evolution of a system, one could let the NN automatically infer the correct evolution law from a small sequence instead of from a single initial condition. Indeed, this idea was exploited in Ref.~\cite{yangPATTERNS2021}, where the solidification latent heat variable was hidden from the CRNN and was implicitly inferred by the model from the provided 10 initial steps.

The opposite possibility of explicitly providing as input the parameters defining the dynamics and a single initial condition, which is closest to a traditional simulation scheme, has also been recently explored. Some works~\cite{fantasiaADVTHESIM2026, bonneville2026towards} have indeed conditioned the NN prediction with external parameters. For example, in Ref.~\cite{fantasiaADVTHESIM2026}, the mismatch between lattice parameters for pure phases was passed to the CRNN to predict the spinodal decomposition in coherent alloys. 

A direct comparison of these two competing approaches using similar architectures, however, is not currently available in the literature. Our work addresses this question and quantitatively compares the prediction performances of CRNN models derived from Refs.~\cite{lanzoniPRM2022, fantasiaADVTHESIM2026} for the case study of crystal growth. Indeed, the acceleration of simulations and composition/process screening offered by ML tools are particularly appealing in this field and interest in these applications has substantially increased in recent years~\cite{kutsukakeJCG2024, petkovicProcCGCM2025, luCGD2024, LuAdvFuncMat2026}. In the case at hand, we consider the prototypical Allen-Cahn dynamics~\cite{allen1979,liCCP2009} in two dimensions. In order to mimic crystal faceting, key for realistic and reliable simulations, we introduce an anisotropic kinetic coefficient producing hexagonal morphologies. Variable values of supersaturation, i.e., of the difference in chemical potential between the gas or liquid mother phase and the crystal itself, are admitted. Hence, since the supersaturation directly controls the actual growth rate of the crystalline phase and the fine-details of faceting, the evolution sequence from a given initial state is parametrically dependent on its value. Although very simple, this model contains the fundamental aspects that characterize the faceted growth of crystals while giving the possibility of constructing large datasets with little computational effort, thus enabling the extensive testing of the NN behaviour. 

The study finds that explicitly conditioning on the supersaturation value reliably increases the quality of predictions and yields more data-efficient NN models. In fact, we find that to obtain the same accuracy, a $\approx 15\times$ larger dataset is required for CRNN exploiting small sequences for implicit inference with respect to explicitly conditioned ones. Based on our findings, this strategy should therefore be regarded as the better option, i.e., explicit information should always be preferred over the automatic correlation discovery capabilities of deep learning methods, whenever possible.

The paper is organized as follows. In Sect.~\ref{sec::methods} we define the Allen-Cahn model with the phase-field approach and provide the technical description of the two NN architectures. Then in Sect.~\ref{sec::results} we analyse in-depth the NN capabilities. First, we analyse the training results (Sect.~\ref{sec::training}). Then, we quantitatively estimate the accuracy of the different trained models on extensive test sets (Sect.~\ref{sec::testing}), inspecting their sensitivity to different supersaturation values. Finally, in Sect.~\ref{sec::largedomain}, we exploit the full potential of our CRNN architecture for large-scale simulations, beyond the training domain size, and in Sect.~\ref{sec::coverage} we analyse the effect on prediction quality due to the density of seeds in the initial conditions.

\section{Methods}\label{sec::methods}

\subsection{Phase-field simulations}\label{sec::pf}

\noindent A simple two-dimensional Allen-Cahn model is considered as a prototypical description of crystal growth under isothermal conditions \cite{provatas_phasefield_2010}. In particular, we consider an order parameter $\varphi$ distinguishing the crystal phase ($\varphi=1$) from its liquid or gaseous mother phase ($\varphi=0$) and set a standard Ginzburg-Landau energy functional $G[\varphi]$ having the two phases as minima, biased by a supersaturation parameter $\Delta\mu$:
\begin{equation}\label{eq::energy}
    G[\varphi]=\int_\Omega \left[\frac{\epsilon}{2}|\nabla\varphi|^2 + w(\varphi) - p(\varphi)\Delta\mu  \right] dx
\end{equation}
with $\epsilon$ the width of the diffuse boundary between the two phases, $\Omega$ the integration domain, and a bulk energy set by the symmetric double-well potential $w(\varphi)=(18/\epsilon) \varphi^2(1-\varphi)^2$ and a smooth, bias function set as $p(\varphi)=\varphi^3(10-15\varphi+6\varphi^2)$.

The growth process is then traced implicitly by the time evolution of $\varphi$ as determined by the local chemical potential $\mu=\delta G/\delta\varphi$ as
\begin{equation}\label{eq::allencahn}                            
    \frac{d\varphi}{dt} = -k(\alpha)\frac{\delta G}{\delta \varphi} = - k(\alpha) \left[-\epsilon \nabla^2\varphi + w'(\varphi) - p'(\varphi)\Delta\mu \right]
\end{equation}
with $k(\alpha)$ a kinetic coefficient, eventually dependent on the local profile orientation $\alpha=\arctan(\nabla\varphi_y/\nabla\varphi_x)$ in order to account for anisotropic growth rates. In the present work, we consider $k(\alpha)=1+\beta\cos(N\alpha)$, with $\beta=0.8$ and $N=6$ to produce hexagonally faceted crystals. For the sake of simplicity, all seeds share the same crystallographic axis thus mimicking epitaxial growth.

Eq.~\ref{eq::allencahn} is here solved numerically by exploiting a simple forward Euler integration scheme and finite differences on a square grid with periodic boundary conditions. A value of $\epsilon=6$ pixels is set and a time step of $0.05/\epsilon$ is used.

\begin{figure}[t]
    \centering
    \includegraphics[width=\linewidth]{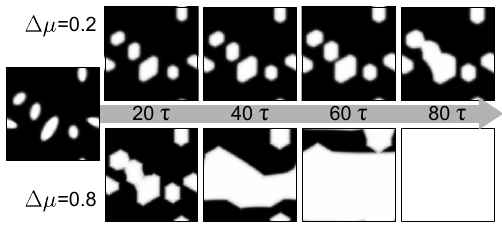}
    \caption{Example of evolution sequences from the same initial configuration for the two limiting values of $\Delta\mu$, i.e. $0.2$ and $0.8$ considered in our training set. Each time unit $\tau$ corresponds to $25$ time steps.}
    \label{fig::sample}
\end{figure}

In Fig.~\ref{fig::sample} we report two evolution sequences starting from the same initial profile and setting a different supersaturation $\Delta\mu$. As made evident by the similarity between the profile obtained for $\Delta\mu=0.2$ at time $80\tau$ and the one for $\Delta\mu=0.8$ at $20\tau$, the main effect of $\Delta\mu$ is to directly control the overall growth rate, providing a relative rescaling of the time-scale. A closer inspection however shows substantial differences in the contours, more rounded for the lowest $\Delta\mu$ and characterized by sharper corners for the highest value, as due to the different weights of the gradient-energy contribution and the double-well imbalance ($\propto\Delta\mu$) to the chemical potential in eq.~\ref{eq::allencahn}.

\subsection{Neural Network approach}

\noindent In this work, the Convolutional Recurrent Neural Network (CRNN) developed in~\cite{lanzoniPRM2022, lanzoniMLST2024} is specialized to reproduce the Allen-Cahn dynamics and adapted to tackle a heterogeneous dataset of time evolution sequences as determined by randomly chosen supersaturation parameters $\Delta\mu$. The CRNN objective is to generate a sequence of subsequent stages during the crystalline material growth, provided one or more snapshots of previous morphologies. In this work, we focus on two variants of the architecture, whose main difference is whether the model is aware of the actual value of the supersaturation parameter.

\begin{figure}[t!]
    \centering
    \includegraphics[width=\linewidth]{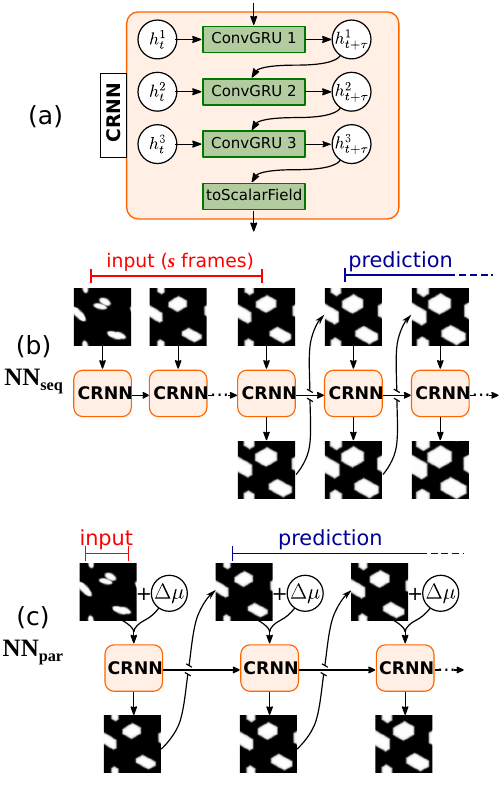}
    \caption{(a) Schematics of the used convolutional recurrent neural network (CRNN) internal architecture. (b,c) \NNseq and \NNpar schematics, showing the difference in the input handling. The internal architecture of the CRNN block is the one reported in panel (b).}
    \label{fig::archiNN}
\end{figure}

In practice, both models are implemented as a convolutional recurrent GRU~\cite{shiARXIV2015, lanzoniPRM2022}, composed of three recurrent stacked layers as sketched in Fig.~\ref{fig::archiNN}(a). Kernel size used is $5\times5$, with circular padding encoding periodic boundary conditions by construction. Hidden states have $16$ channels, for a total of about $140000$ parameters for both models. The next state of the system $\varphi_{t+\tau}$ is obtained by summing the output of the NNs to the previous phase-field map $\varphi_t$, i.e., the models are required to learn the residual between subsequent states of the system. At variance with the work in Ref.~\cite{lanzoniMLST2024}, no mass conservation is required in the current case, hence the NN output does not need any additional processing. A schematic for both architectures is shown in Fig.~\ref{fig::archiNN}(b) and Fig.~\ref{fig::archiNN}(c).

The first NN, which will be called \NNseq in the following, does not receive $\Delta \mu$ as an explicit parameter, but rather is required to infer it by processing as input a mini-sequence of $s$ frames. This is made possible by the recurrent layers in the NN architecture, which act as a sort of "memory" capable of implicitly extracting temporal information from input sequences. While this approach is particularly appealing for applications where the driving forces for the dynamics are not known or not completely specified (e.g., in experimental contexts), it also has the downside of always requiring multiple time-frames as input for the NN model. In this sense, the goal of \NNseq could be understood as a sequence-completion task.

On the other hand, the second model \NNpar is explicitly provided the actual value of the supersaturation $\Delta \mu$. To this aim, the numerical value of the supersaturation is converted into a uniform spatial tensor and concatenated to the $\varphi$ map for the current timestep, using the same strategy as in~\cite{fantasiaADVTHESIM2026}. The input of \NNpar model is therefore composed of a two-channel image, with the first channel containing the material morphology as described by the phase field $\varphi$ and the second being a constant map equal to $\Delta \mu$. This simple but effective encoding scheme can, in principle, be straightforwardly extended to non-constant driving forces. Once $\Delta \mu$ and the initial condition are explicitly known, the full evolution of the system is completely determined, since the underlying equation of motion Eq.~\ref{eq::allencahn} is first-order in time. For this reason, \NNpar is provided with a single snapshot to predict the full growth.

\section{Results and discussion}\label{sec::results}

\subsection{Training and Validation}\label{sec::training}

\noindent A dataset of $7500$ PF simulation sequences is constructed for the training of both NN models. Each case consists of a series of $200$ time-frames (taken every $\tau=25$ integration time-steps) reporting the $\varphi$ field as evolving from a random distribution of elliptical seeds of different eccentricity and orientation toward complete coverage of the domain. A $128 \times 128$ domain is considered and each grid value $\varphi(i,j)\in[0,1]$ defines a pixel in the gray-scale frame image. The value of supersaturation $\Delta\mu$ is uniformly sampled within the range $[0.2, 0.8]$, corresponding to a $\approx4\times$ variation in the net growth rates (see Fig.~\ref{fig::sample}) and encompassing both rounded and sharp facet corners.

A set of $N_\text{set}$ sequences is extracted from the full dataset for the training of the NN, with a $4:1$ random partitioning between training and validation. For each sample, only a segment of $T=50$ contiguous time frames is selected at random to expose the NN to diverse evolution stages.

The NN training is performed using the standard implementation of the Adam optimizer \cite{kingma2017}. The loss function $\mathcal{L}$ is defined as the common mean squared error (MSE) between the PF ground-truth ($\varphi^\text{PF}$) and NN-predicted ($\varphi^\text{NN}$) profiles, averaged on all time-frames $t$ of each sequence and on all $N_\text{b}$ samples in the training batch:
\begin{equation} \label{eq::loss}
    \mathcal{L} (\vartheta) = \frac{1}{N_{\text{b}}T} \sum_{n=1}^{N_{\text{ts}}} \sum_{t=1}^{T} \langle\left[\varphi^\text{PF}_n(t) - \varphi^{\text{NN}}_n(t|\vartheta) \right]^2 \rangle
\end{equation} 
with $\vartheta$ the set of NN parameters and $\langle \, . \, \rangle$ indicates the spatial average. The same loss function is also computed on the validation set. Mirror symmetry is taken into account as in Refs.~\cite{lanzoniPRM2022, lanzoniMLST2024} via data augmentation. A mini-batch size of $N_\text{b} = 3$ has been used.

\begin{table}[b]
    \centering
    \begin{tabular}{| c | c | c c|}
        \hline
         arch. & ID & $N_\text{set}$ & $s$ \\
         \hline
         \multirow{5}{*}\NNseq & \NNs{5}{1.5k} & 1500 & 5 \\
         & \NNs{5}{5k} &  5000 & 5 \\
         & \NNs{5}{7.5k} & 7500 & 5 \\
         & \NNs{3}{5k} & 5000 & 3 \\
         & \NNs{7}{5k} & 5000 & 7 \\
         \hline
         \multirow{3}{*}\NNpar & \NNp{0.5k} & 500 & 1 \\
         & \NNp{1.5k} & 1500 & 1 \\
         & \NNp{5k} & 5000 & 1 \\
         \hline
    \end{tabular}
    \caption{List of NN models investigated in the present study for both \NNseq and \NNpar architectures, labelled according to dataset size $N_\text{set}$ and length $s$ of the input sequence.}
    \label{tab::models}
\end{table}

Given the recurrent architecture, the curriculum learning technique \cite{bengioPROC2009} is implemented, starting from the prediction of only the last frame of a sequence in the first training epoch and then gradually reducing the number of input frames while requesting the prediction of the increasing number of remaining sequence steps, down to the actual mini-sequence of length $s$ for \NNseq or single frame for \NNpar decided as input. During the process, the loss is evaluated on the predicted frames. While resulting in an apparent raise of the loss, this gradual increment of the complexity of the NN task has been demonstrated to return more stable and efficient convergence of the model.

For the present work, different models have been trained by varying both the size of the dataset $N_\text{set}$ and, for \NNseq, the length $s$ of the input mini-sequence, as listed in Table~\ref{tab::models}.

\begin{figure}[b]
    \centering
    \includegraphics[width=\linewidth]{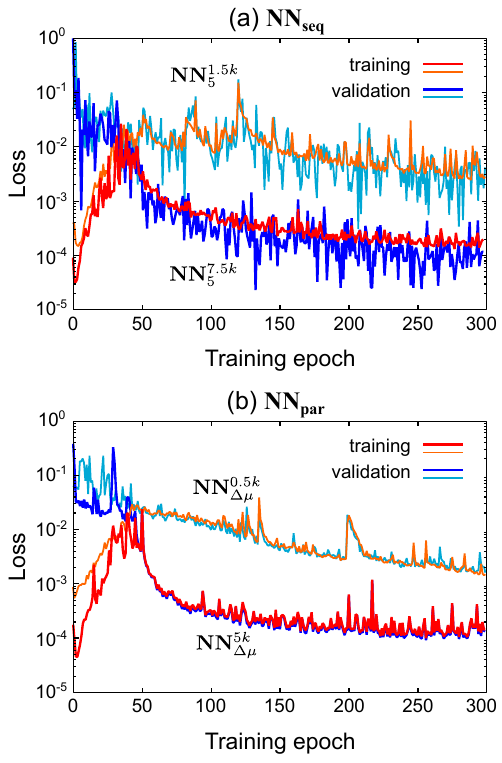}
    \caption{Training and validation losses during training for the mini-sequence model \NNseq (a) and the parameter model \NNpar (b). The cases for smaller and larger dataset sizes are shown for both.}
    \label{fig::lossplot}
\end{figure}

\begin{figure*}[t!]
    \centering
    \includegraphics[width=0.9\linewidth]{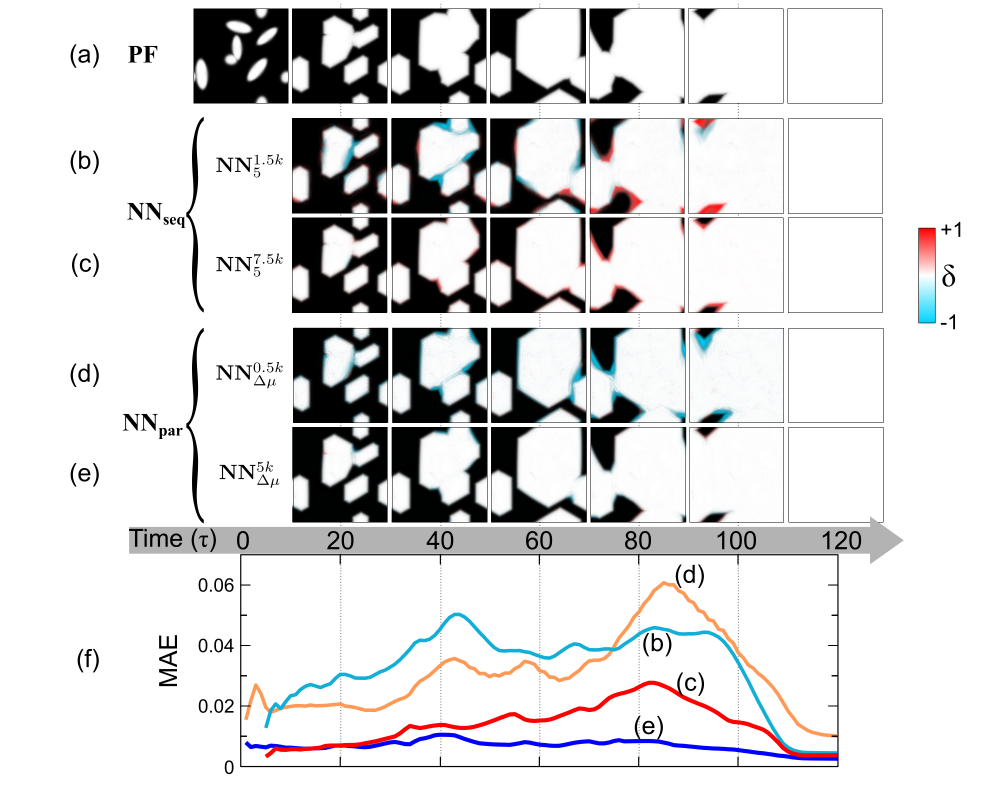}
    \caption{Evolution sequence for a representative case ($\Delta\mu=0.6)$ in the test set comparing the ground-truth PF solution (a) with the corresponding predictions from both \NNseq (b,c) and \NNpar (d,e). The pixel-by-pixel error $\delta$ is superimposed on the predicted images by colour bar. (f) Time evolution of the MAE between NN predictions and the true PF solution.}
    \label{fig::compare06}
\end{figure*}

All models showed convergence of the training procedure, with the training and validation losses steadily decreasing, albeit for fluctuations introduced by the stochastic Adam algorithm, with no sign of overfitting. In Fig.~\ref{fig::lossplot}, we report the evolution of the training and validation losses obtained over $300$ training epochs for the models with larger and smaller datasets for both \NNseq (a) and \NNpar (b). Reasonably, the models trained on the largest datasets exhibit a faster decay rate of the loss for each epoch, leading to a final value about one order of magnitude lower than those trained on the smallest. This trend is consistently observed also for the intermediate models \NNs{5}{5k} and \NNp{1.5k} (see Fig. S1 of Supplementary Material). The effect of considering a different mini-sequence length $s$ is instead less evident, but slightly lower loss values for longer input sequences may be observed (see Fig. S2 of Supplementary Material).
Notably, \NNs{5}{7.5k} and \NNp{5k} return similar loss values despite the different architecture and dataset size, suggesting similar prediction performances. This expectation will, however, be confuted by the extensive testing in the following sections.

After each training run, the model with the lowest validation loss within the last $50$ epochs is selected as the best performing one and used for all the following analyses.

\subsection{Testing the NN prediction performances}\label{sec::testing}

\begin{figure*}[t]
    \centering
    \includegraphics[width=0.9\linewidth]{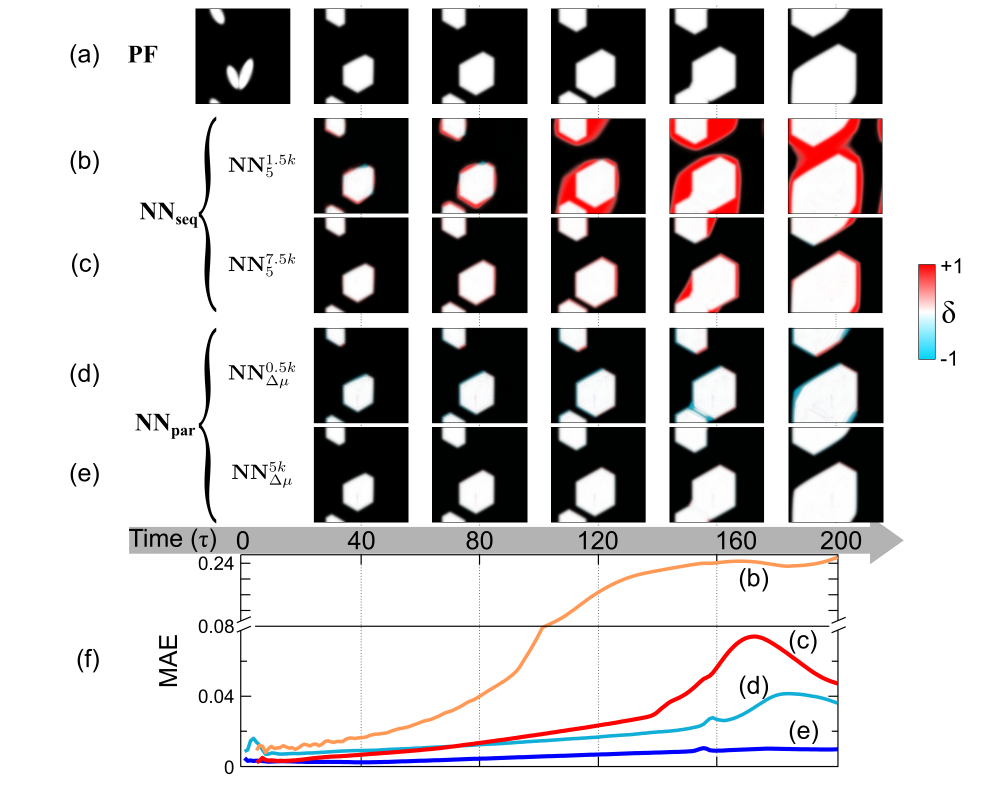}
    \caption{Evolution sequence for a representative case ($\Delta\mu=0.3)$ in the test set comparing the ground-truth PF solution (a) with the corresponding predictions from both \NNseq (b,c) and \NNpar (d,e). The pixel-by-pixel error $\delta$ is superimposed on the predicted images by colour bar. (f) Time evolution of the MAE between NN predictions and the true PF solution.}
    \label{fig::compare03}
\end{figure*}

\noindent In order to evaluate the predictive performances of the trained models, an independent test set of $2500$ simulations is generated with the same criteria used for the training and validation sets. All models in Table~\ref{tab::models} are then used to predict the full $200\tau$-long time evolution sequence for all these cases to assess their performance. Notice that this task exceeds by $4\times$ the $50$-frame-long sequences to which the NNs are exposed during training, thus serving as a generalization test to longer simulation times.

As a first evaluation, predicted frames can be "visually" compared with the corresponding ground-truth ones computed from the numerical integration of the PF eq.~\eqref{eq::allencahn}. In Fig.~\ref{fig::compare06} and Fig.~\ref{fig::compare03} we consider two representative evolutions for different values of $\Delta\mu$ ($0.6$ and $0.3$ respectively) and compare the predictions for both \NNseq and \NNpar, again focusing on the models trained on the smallest and largest datasets so to appreciate the best and worst performance for both classes (see Fig. S3 and Fig. S4 of Supplementary Material for the analyses on all other NN models). In panels (b-e) of both figures, the pixel-by-pixel signed difference between the NN-predicted $\varphi$ field and the PF one, i.e. $\delta=\varphi^\text{NN}-\varphi^\text{PF}$, is computed and superimposed by colour map to the black-and-white output images so to evidence the regions where the predictions are in excess ($\delta>0$) or defect ($\delta<0$) with respect to the true ones (a).

For a more quantitative estimation of the prediction errors, we also monitored the Mean Absolute error (MAE) defined as the domain average of the absolute value of $\delta$, i.e., MAE=$\langle|\delta|\rangle$, as a function of time. Since the $\varphi$ field is bounded in the range $[0,1]$, the MAE value can be directly taken as the fraction of mispredicted pixels in the image. The time evolutions of MAE by the different NN for both evolutions in Fig.~\ref{fig::compare06} and Fig.~\ref{fig::compare03} (see also Fig. S3(f) and Fig. S4(f)) are reported in the respective panels (f).

In the case of Fig.~\ref{fig::compare06}, we observe a relatively fast growth leading to full domain coverage after about $120\tau$. We can observe how the evolution sequences are well captured by all models. Most discrepancies are localized in the regions where crystalline units coalesce, i.e., where the dynamics abruptly change because of the change in topology, causing a slight delay or speed up that does not significantly impact the overall evolution. The largest errors, still leading to a MAE always less than $0.06$ are found for the models trained on smaller datasets \NNs{5}{1.5k} and \NNp{0.5k}, while the best performance is achieved by \NNp{5k}, which is always below $0.01$ MAE. It can be noticed how, as time advances, the MAE generally grows because of error accumulation, resulting in steeper peaks when coalesce events occur and returning to zero when the domain is fully covered.

In the case of Fig.~\ref{fig::compare03}, characterized by a slower dynamics leading just to a partial filling of the domain, we instead observe a neat difference in the prediction performance with the \NNseq models struggling to advance the profile at the right rate in contrast to the \NNpar models which are still providing a good reproduction of the sequence but for local errors. While \NNs{5}{1.5k} results are completely unreliable as tracing a constantly faster dynamics that lead to MAE accumulation beyond $0.2$, the prediction given by \NNs{5}{7.5k}, although more reasonable, returns a MAE error that peaks two times larger than the one of \NNp{0.5k}. Both \NNp{0.5k} and \NNp{5k} are in line with the MAE errors found in the previous case of Fig.~\ref{fig::compare06}. Notably, the \NNp{0.5k} model reports a MAE error below \NNs{5}{7.5k} for a large part of the evolution.

\begin{figure}[t!]
    \centering
\includegraphics[width=\linewidth]{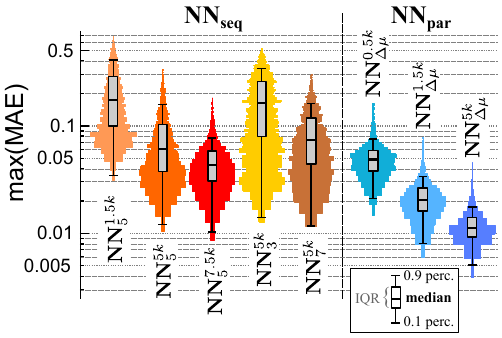}
    \caption{Distribution of the maximum of MAE for the predictions of time evolutions in a test set of $2500$ sequences conformal to the training dataset for the different NN models. The full distribution is shown in colour by violin plots. The boxes correspond to the interquartile range (IQR). The median value and the $[0.1,0.9]$ percentile range are also shown.}
    \label{fig::histo}
\end{figure}

Although specific of the selected simulation cases, the analysis referred to Fig.~\ref{fig::compare06} and Fig.~\ref{fig::compare03} is consistent for all simulations composing the test set. Each case indeed results in an error accumulation in the early growth stages, reaching a maximum at intermediate coverage and then going back to zero at later times. This error accumulation becomes critical for \NNseq models when considering low supersaturations.

To better quantify the overall accuracy of the NN prediction across the entire test set, we now inspect the statistical distribution of the maximum MAE value for each evolution sequence as obtained from any NN model in Table~\ref{tab::models}. It must be noted that this analysis provides the worst-case scenario for the prediction error, as within the time sequence, the discrepancy is generally lower than that. The results are reported in Fig.~\ref{fig::histo}, showing the error distributions as violin plots along with the corresponding box-plot representation tracing the median and interquartile range as well as the $[0.1,0.9]$ percentile range. As already noted from the loss comparisons in Fig.~\ref{fig::lossplot} and by the sequence comparisons in Fig.~\ref{fig::compare06} and Fig.~\ref{fig::compare03}, the size of the training dataset is crucial to achieve better accuracy. 

In the case of the \NNseq architecture, only the largest dataset of $7500$ sequences guarantees that more than 90\% of the predictions yield max MAE errors below an acceptable $0.075$ threshold, with a median value of $0.043$. By comparing the \NNseq models differing just in the mini-sequence length $s$, we clearly see that the choice of $s=5$ is the optimal one. Indeed, for shorter sequence \NNs{3}{5k} the NN performs poorly with a broad error distribution with median error almost double the one of \NNs{5}{5k}, probably due to the insufficient information to infer the proper evolution rate. On the other hand, the \NNs{7}{5k} model, taking a 2-frame longer initial sequence than \NNs{5}{5k} does not show any benefit from it, exhibiting almost the same error distribution at the price of more input frames.

\NNpar models perform significantly better than the \NNseq ones, to the point that the worse of the firsts, \NNp{0.5k}, is on par with the best among the seconds, i.e. \NNs{5}{7.5k}, yielding 90\% of the predictions beyond the same $0.075$ error threshold and a median value of $0.048$, despite being trained on a $15$ times smaller dataset and corroborating the superior data efficiency of this approach. The best performer among our tests is \NNp{5k}, yielding more than 90\% of the predictions with errors below $0.018$ and a median value of $0.011$.

The supremacy of \NNpar was expected as the explicit conditioning by supplying the correct $\Delta\mu$ value as input strongly simplifies the task with respect to inferring it implicitly from a short mini-sequence, as requested to \NNseq models. The similarity between \NNp{0.5k} and \NNs{5}{7.5k} indicates that the performance gap of \NNseq can be closed only at the price of a more demanding training on significantly larger datasets. It is also worth to point out that the difference between \NNpar and \NNseq revealed by this testing was not apparent from the training and validation losses in Fig.~\ref{fig::lossplot}, thus underlying the importance of independent testing.

\subsection{NN performances as a function of supersaturation}\label{sec::deltamu}

\begin{figure}[t!]
    \centering
    \includegraphics[width=\linewidth]{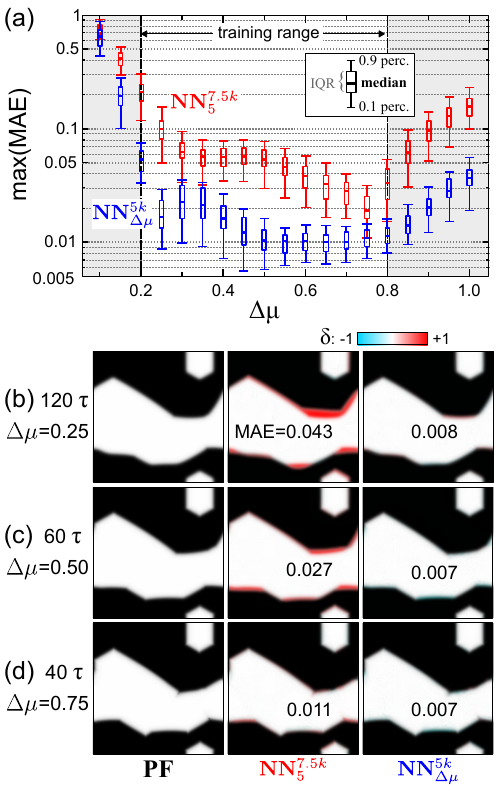}
    \caption{(a) Dependency of the maximum of MAE distribution on the supersaturation value $\Delta\mu$ for both \NNs{5}{7.5k} and \NNp{5k}. Each box represents the interquartile range (IQR) of the error distribution on a dataset of $100$ sequences starting from the same initial profiles and evolving according to the corresponding supersaturation $\Delta\mu$ up to complete filling of the domain. The median value and the $[0.1,0.9]$ percentile range are also shown. The shaded areas represent $\Delta\mu$ values outside the range of the training set. (b-d) Comparison between true PF and NN-predicted profiles, for an equivalent growth stage from the same initial configuration of Fig.~\ref{fig::sample} and different $\Delta\mu$. The pixel-by-pixel error $\delta$ is superimposed on the predicted images by colour bar along with the value of MAE.}
    \label{fig::mae_dmu}
\end{figure}

\noindent As suggested by the comparison of Fig.~\ref{fig::compare06} and Fig.~\ref{fig::compare03}, the NN models, especially \NNseq ones, could behave differently according to the actual value of $\Delta\mu$. We here inspect a $[0.1,1.0]$ range of $\Delta\mu$ values, slightly exceeding the one of the training set, evaluating the NN extrapolation capabilities too. To this purpose, we prepared new test sets for each value of $\Delta\mu$, each composed of $100$ sequences. To avoid biases due to the initial random configuration, the same $100$ initial profiles are adopted for all test sets. Moreover, we do not fix the number of frames of the test sequence but we let the simulations run until reaching complete filling of the domain to determine the actual maximum of MAE on the complete dynamics. Domain filling is simply evaluated though the coverage $\theta $ defined as the average value of $\varphi$ over the whole domain. The complete filling condition is defined by $\theta > 0.999$. Following this procedure, the length of the test sequences changes from less than $100\tau$ for the highest $\Delta\mu$ values to more than $800\tau$ for the lowest $\Delta\mu$, thus providing in this latter case a further test of time extrapolation. The results of this analysis, for the best-performing models in the two classes, i.e. \NNs{5}{7.5k} and \NNp{5k}, are illustrated in Fig.~\ref{fig::mae_dmu}, reporting for each sampled $\Delta\mu$ a box-plot showing the median value, interquartile range and $[0.1,0.9]$ percentile range of the distribution of MAE maxima (see also Fig. S5 of Supplementary Material for the same analysis on the original test set, for all trained models). 

It can be seen that the \NNp{5k} outperforms \NNs{5}{7.5k} for any $\Delta\mu$, with maximum MAE errors which remains well below $0.04$ for 90\% of the test cases for all $\Delta\mu$ values within the training range. A pronounced increase in the error is, however, observed when extrapolating for $\Delta\mu\le0.2$ with predictions becoming unreliable right below that threshold. Extrapolation toward higher $\Delta\mu$ beyond $0.8$ is instead possible, even if the error increases sharply, thus limiting the accessible range to $\Delta\mu\le1.0$. 

The behaviour of maximum MAE errors for \NNs{5}{7.5k} instead shows a more pronounced dependence on $\Delta\mu$ also within the training range. The best results are found when approaching the higher bound of $\Delta\mu = 0.8$. At intermediate values, the error tends to stabilize below a threshold of $0.08$ for 90\% of the test cases, similarly to the overall values found from Fig.~\ref{fig::histo}. However, when decreasing $\Delta\mu$ below $0.3$, the error rapidly increases so that a large fraction of the NN predictions within the lowest $\Delta\mu$ portion of the training range are actually subject to large MAE beyond $0.1$. The model also fails in extrapolating beyond the training range as errors quickly increase. By this analysis, we conclude that the \NNs{5}{7.5k} should only be trusted for predictions within the range of $\Delta\mu\in[0.3,0.8]$ despite the training including also lower values. A possible interpretation for this failure at low $\Delta\mu$ may be related to the fact that such low values imply a very small profile change between the few frames in the mini-sequence, which limits the NN model's capability of recognizing the actual evolution rate.

As from Sect.~\ref{sec::pf} it was noted how a change in $\Delta\mu$ affects the smoothness of the growth front morphology, beyond a mere rescaling of the growth rate. Assessing the trained models' ability to reproduce such fine details is the ultimate test of their robustness. In Fig.~\ref{fig::mae_dmu}(b-d) we compare the profiles obtained by simulations started from the same initial configuration of Fig.~\ref{fig::sample} for three different values of $\Delta\mu=0.25$, $0.50$ and $0.75$ at times $120 \tau$, $60 \tau$ and $40 \tau$ scaled to return equivalent growth stages. As evident from the PF ground-truth profiles, the morphologies are similar, except for the aforementioned trend toward sharper edges with increasing $\Delta\mu$. Both \NNs{5}{7.5k} and \NNp{5k} models return an overall satisfactory reproduction of such profiles. Significant local discrepancies are only evident for \NNs{5}{7.5k} at low $\Delta\mu$, in line with the previous analysis. At a close inspection it can also be appreciated how the predicted profiles are characterized by rounder corners for low $\Delta\mu$ while sharper edges are obtained for larger values (see also Fig. S6 of Supplementary Material), indicating that the trained models fully learned the role of the $\Delta\mu$ parameter in controlling also the finer details in the profile evolutions. It is particularly remarkable that MAE prediction errors for \NNp{5k} and for \NNs{5}{7.5k} at sufficiently large $\Delta\mu$ are typically lower than the discrepancies between the different PF profiles at same evolution stages (typical $\langle|\delta|\rangle\approx0.02-0.04$), thus providing a further proof that the trained models perform better than rescaling over time the same averaged features.

\subsection{Generalization to large domains}\label{sec::largedomain}
\begin{figure*}[t]
    \centering
    \includegraphics[width=\linewidth]{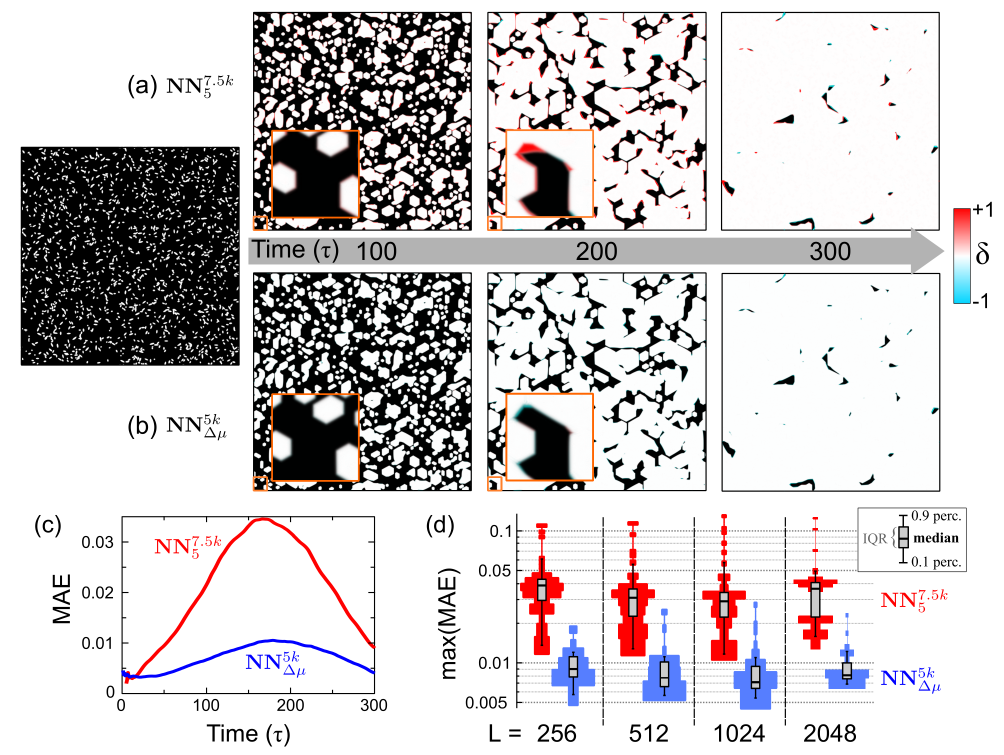}
    \caption{Evolution sequence for crystal growth under the same conditions of training set but on a $2048 \times 2048$ domain ($\Delta\mu = 0.318$) as obtained by models (a) \NNs{5}{7.5k} and (b) \NNp{5k}. Pixel-by-pixel errors $\delta$ are superimposed on the evolution frames by colour map. Insets provide a magnification of the $128\times128$ left-bottom corner. (c) Time evolution of MAE or the sequences in (a) and (b). (d) Analysis of the distribution of MAE maxima across $50$ evolution sequences for different domain sizes $L$ by both violin plot and box plots showing the median, interquartile range (IQR), and $[0.1, 0.9]$ percentile range.}
    \label{fig::size}
\end{figure*}

\noindent Both NN approaches discussed here exploit a fully-convolutional architecture so that the trained models can be, in principle, applied to any domain size, eventually much larger than the one considered for its training. This is particularly appealing in the perspective application of this class of models, since it would allow training on small-size, low-cost simulation domains, while promising the exploitation for larger domains at a fraction of the computational cost that using PF would imply.

To prove this potential, we then analysed the NN performances in predicting the evolution sequence for $50$ test cases defined with the same conditions of the training dataset but extending on $L\times L$ domain sizes of $256 \times 256$, $512 \times 512$, $1024 \times 1024$ and $2048 \times 2048$ collocation points.

An example of NN predicted profiles on the largest $2048 \times 2048$ domain is reported in Fig.~\ref{fig::size} for the \NNs{5}{7.5k} (a) and \NNp{5k} (b) best models (see Fig. S7 for additional examples on other domain sizes). The simulation corresponds to a value of $\Delta\mu=0.318$ requiring a time of about $300\tau$ to reach complete filling of the domain, i.e., $6$ times the duration of the training sequence. A close inspection of the pixel-by-pixel error $\delta$, highlighted in the insets, clearly shows local discrepancies in the \NNs{5}{7.5k} predicted profiles, while for \NNp{5k} they are barely distinguishable. The evaluation of the MAE during the time evolution, reported in Fig.~\ref{fig::size}(c), shows a progressive error accumulation reaching a maximum around $t=170\tau$, when most of the largest domains coalesce, and then decreasing to zero as the domain is fully covered. The MAE peak values are as small as $0.035$ for \NNs{5}{7.5k} and $0.010$ for \NNp{5k}.

\begin{figure*}[t]
    \centering
    \includegraphics[width=\linewidth]{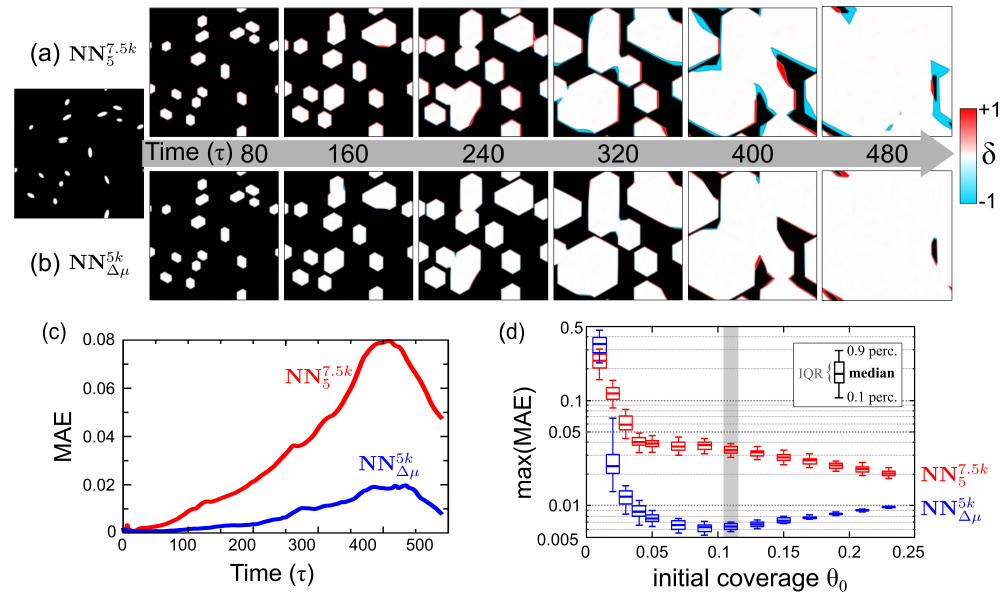}
    \caption{Evolution sequence for crystal growth starting from a low coverage configuration of $\theta_0=0.023$ on a $512 \times 512$ domain, with $\Delta\mu=0.5$, as obtained by the models (a) \NNs{5}{7.5k} and (b) \NNp{5k}. Pixel-wise errors $\delta$ are superimposed on the evolution frames by colour map. (c) Time evolution of MAE for the sequences in (a) and (b). (d) Dependence of the MAE maxima distribution as a function of the initial profile coverage. Box-plots show the median, interquartile range (IQR) and $[0.1,0.9]$ percentile range for $50$ evolution sequences with $\Delta\mu=0.5$. The gray shaded area corresponds to the training set coverage range.}
    \label{fig::cov}
\end{figure*}

While the reported case of Fig,~\ref{fig::size} can be taken as representative of the test set simulations, a complete quantitative analysis of the NN accuracy can only be achieved by inspecting the actual distribution of MAE maxima across the test case, as already done for the original $128 \times 128$ domain size in Fig.~\ref{fig::histo}. In Fig.~\ref{fig::size}(d), we report by violin plots and box plots the resulting distributions for the tests on the different domain sizes, still comparing \NNs{5}{7.5k} and \NNp{5k}. We can see that the errors are consistent across all sizes, with the \NNs{5}{7.5k} capable of returning 90\% of the predictions with maximum MAE errors below $0.06$ and the \NNp{5k} remaining just around $0.01$. We then conclude that the NN can reliably be extended to larger domain sizes, ideally approaching the experimental ones, with no loss in accuracy with respect to the original training conditions.

\subsection{Initial coverage effect}
\label{sec::coverage}

\noindent A last key aspect that has been neglected so far in our tests is the role of the different initial distribution of seeds. All the previous analysis indeed applied the same protocol to generate the initial profiles, returning fairly consistent seed distributions that correspond to an initial coverage $\theta$ of $0.11\pm 0.015$. 

However, when considering larger domains, it becomes natural to consider different seed densities, in particular enabling the analysis of more sparse configurations. In this last section, we inspect how the trained NN can possibly generalize with respect to variable initial coverages. In Fig.~\ref{fig::cov} we consider the time evolution in a $512 \times 512$ domain as an example, initiated from a seed distribution corresponding to a coverage of about $\theta_0=0.023$, i.e., about $5$ times lower than the ones of training. The supersaturation condition is set to $\Delta\mu=0.5$.  The complete evolution up to full domain filling extends for about $500\tau$, so that the comparison extends $10$ times beyond the training sequence duration. In Fig.~\ref{fig::cov}(a), we have the evolution predicted by the \NNs{5}{7.5k} model, while in panel (b) the one from \NNp{5k} is reported. The pixel-wise errors in the predictions are highlighted by the $\delta$ colour map, while the evolution of MAE is shown in Fig.~\ref{fig::cov}(c). It can be observed that both models return quite consistent evolutions despite the unusual initial configuration with respect to training conditions. Consistent with the rest of the study, \NNp{5k} performs better, returning a maximum MAE of just $0.02$ while \NNs{5}{7.5k} returns larger errors at merging domains, reaching a still acceptable MAE of $0.08$.

To fully characterize the relationship between NN accuracy and initial surface coverage, we generated a series of test sets composed of $50$ cases and initialized with prescribed coverage $\theta_0$ both below and above the typical value used in the NN training. A fixed $\Delta\mu=0.5$ is here considered. For each $\theta_0$ test set, we then analyse the distribution of MAE maxima and report the corresponding median, interquartile range, and $[0.1,0.9]$ percentile range by box-plots in Fig.~\ref{fig::cov}(d). Once again, the best performing models \NNs{5}{7.5k} and \NNp{5k} are compared. Consistent with the previous analyses, \NNp{5k} proves to be the most accurate, returning MAE maxima as low as $0.02$ for more than 90\% of the test cases above a $\theta_0=0.03$ threshold. Predictions are still reliable for $\theta_0=0.02$ but fail for lower coverage. Similarly, also \NNs{5}{7.5k} performs well for all coverages above the $\theta_0=0.03$ threshold but becomes unreliable below that.

We find that in both cases the NN can reliably predict the evolution of profiles starting from initial coverage as low as $0.03$. This holds true for different $\Delta\mu$ values within the range of low-prediction error (i.e., $[0.3,0.9]$; $\Delta\mu=0.3$ and $0.7$ are reported in Fig.~S8 of Supplementary Material).

\section{Conclusions} \label{sec:conclusions}
Two Convolutional Recurrent Neural Network architectures have been analysed in depth, demonstrating their ability to approximate the numerical solution of the Allen-Cahn equation for crystal growth. Notably, the NN surrogates are found to properly recognize the fine effects of variable supersaturation values on the faceted growth morphologies, beyond a mere rescaling of growth rates on averaged features, thus proving their suitability for applications to more realistic and detailed models of specific crystalline materials.

As expected, explicit conditioning by supplying the critical parameter as input, i.e. supersaturation in the present case, ensures the best predictions, yielding the lowest mean absolute error in the cases analysed. Our best model (\NNp{5k}) is found capable of producing quantitative predictions of evolution sequences on the whole spectrum of tested conditions, with local errors limited to just a few pixels (more than 90\% of the tested cases are predicted with less than 1.8\% of wrong pixels). Notably, we find that even when training on $10$ times smaller dataset (model \NNp{0.5k}) the predictions are still generally acceptable (the prediction error grows up to 7.5\%), indicating that the approach can still be usable even if training data were scarcely available.

On the other hand, the \NNseq architecture, inferring the supersaturation implicitly from short sequences, generally requires larger training datasets to produce consistent levels of prediction accuracy. In particular, our analysis suggests that an increment in the dataset size of at least a factor $15$ should be needed to match performances. Despite this higher training cost, this fully data-driven approach could still be the only choice whenever tackling a problem for which the leading parameters are unknown, e.g. dealing with experimental data. Fortunately, the model shows consistent performance already for a mini-sequence as short as $5$ frames, limiting the effort needed to produce a suitable input.

The trained models were also successfully applied to larger computational domains, with no loss of accuracy. Furthermore, they also provided good generalization capabilities for moderate variations in the initial crystal seed density, and for simulation more than $10$ times longer than the training ones, thus making it possible to apply the approach even to more realistic scales.

It is worth noting that the numerical solution of the present Allen-Cahn does not pose severe computational bottlenecks. Indeed, the execution times for the finite-difference algorithm are generally faster than the NN evaluation on the same CPU (here, Intel Core i7-12600k). A significant benefit of a factor $\approx 10 \times$ is achieved only if performing the NN evaluation on a GPU (NVIDIA RTX 3060). Indeed, in this work, the main goal was not to achieve acceleration, but took advantage of the inexpensive model to perform extensive testing. On the other hand, we expect substantial speed-ups offered by the NN surrogate whenever considering more complex, nonlinear dynamics eventually requiring more advanced integration schemes and more costly techniques, e.g., finite element method.

\section*{Data availability statement}
\noindent The datasets used to train and test the model are openly available in Materials Cloud Archive at \href{https://doi.org/10.24435/materialscloud:yv-sy}{https://doi.org/10.24435/materialscloud:yv-sy}, reference number 2026.85. The code used to train the NN model is freely available on GitHub at \href{https://github.com/dlanzo/CRANE}{https://github.com/dlanzo/CRANE}.

\section*{Acknowledgments}
\noindent FM, RB, and DL acknowledge financial support from ICSC—Centro Nazionale di Ricerca in High-Performance Computing, Big Data and Quantum Computing, funded by the European Union—NextGenerationEU. D.L. acknowledges financial support from ICSC SPOKE 7 CNR, project INNOVATOR, CUP B93C22000620006, CN00000013.

\bibliography{biblio}

\noindent 

\clearpage

\makeatletter

\def\maketitle{
\@author@finish
\title@column\titleblock@produce
\suppressfloats[t]}
\makeatother

\setcounter{equation}{0}
\setcounter{figure}{0}
\setcounter{table}{0}
\setcounter{page}{1}
\setcounter{section}{0}
\makeatletter
\long\def\MaketitleBox{%
  \resetTitleCounters
  \def\baselinestretch{1}%
  \begin{center}%
   \def\baselinestretch{1}%
    \Large\@title\par\vskip18pt
    \normalsize\elsauthors\par\vskip10pt
    \footnotesize\itshape\elsaddress\par\vskip36pt
    \end{center}
  }
\makeatother

\title{\hrule\vspace{1cm} SUPPLEMENTARY MATERIAL \\\vspace{1cm}\hrule\vskip0pt\vspace{1.cm} Neural surrogates for crystal growth dynamics with variable supersaturation: \\explicit vs. implicit conditioning}

\maketitle
\onecolumngrid

\renewcommand{\thefigure}{S\arabic{figure}}
\renewcommand{\thesection}{S\arabic{section}}

\vspace{-0.3cm}
\begin{figure}[ht!]
    \centering
    \includegraphics[width=0.9\textwidth]{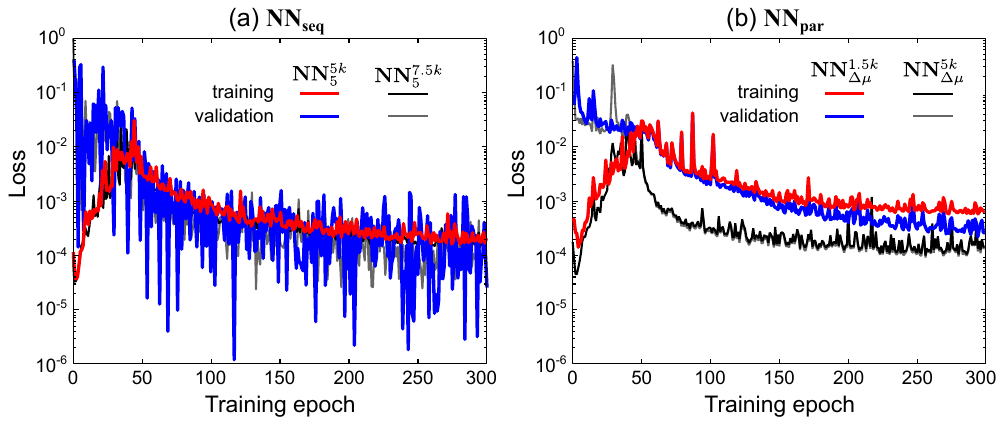}
    \caption{Training and validation losses during training for the mini-sequence model \NNs{5}{5k} vs. \NNs{5}{7.5k} (a) and the parameter model \NNp{1.5k} vs. \NNp{5k} (b).}
    \label{fig::S1}
\end{figure}

\vspace{-0.3cm}
\begin{figure}[h!]
    \centering
    \includegraphics[width=0.45\textwidth]{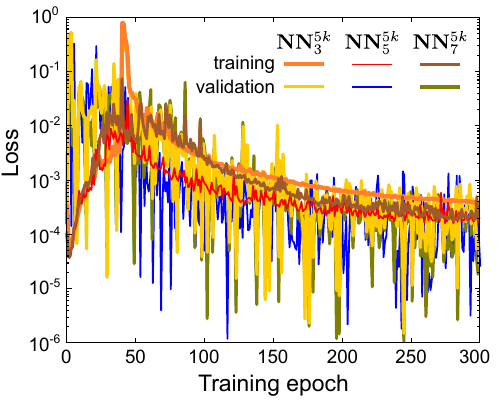}
    \caption{Comparison of the training and validation losses during training for the \NNseq mini-sequence model of 3 (\NNs{3}{5k}), 5 (\NNs{5}{5k}) and 7 (\NNs{7}{5k}) input frames.}
    \label{fig::S2}
\end{figure}

\begin{figure}[ht!]
    \centering
     \includegraphics[width=\textwidth]{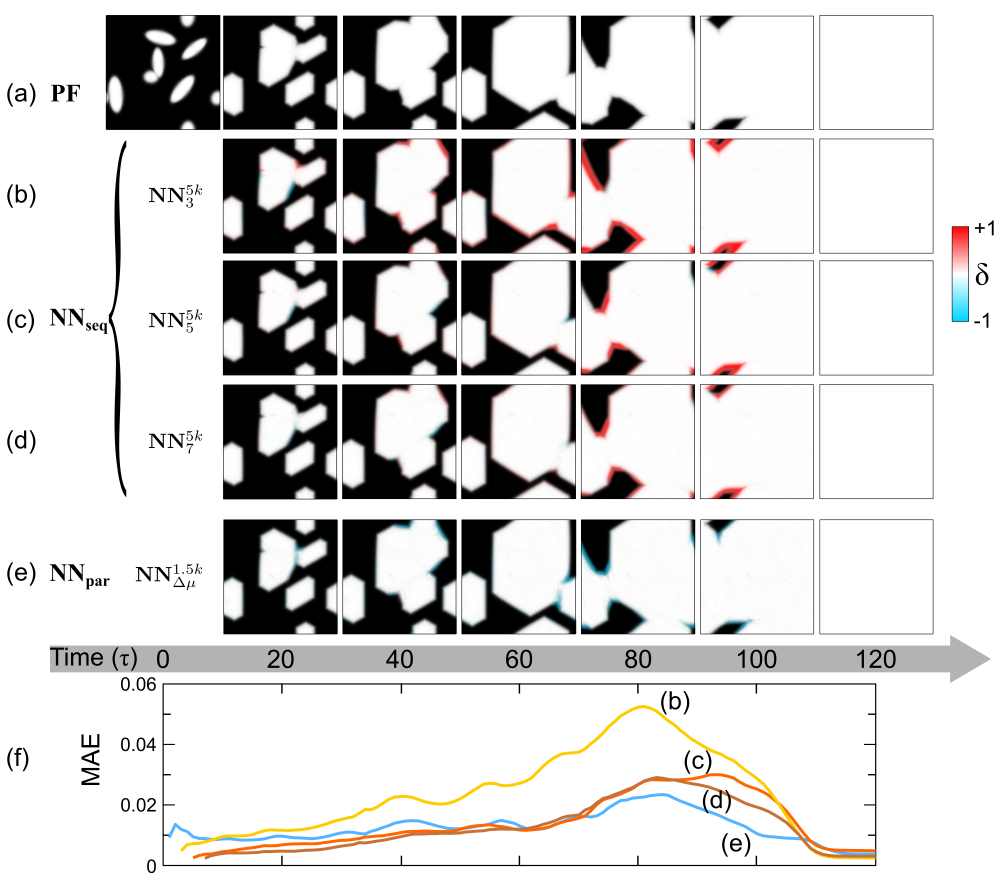}
    \caption{Evolution sequence for the same case of Fig.~4 of the main manuscript ($\Delta\mu=0.6)$ comparing the ground-truth PF solution (a) with the corresponding predictions from both \NNseq (b,c,d) and \NNpar (e) models. The pixel-by-pixel error $\delta$ is superimposed on the predicted images by colour bar. (f) Time evolution of the MAE between NN predictions and true PF solution.}
    \label{fig::S3}
\end{figure}

\begin{figure}[ht!]
    \centering
    \includegraphics[width=\textwidth]{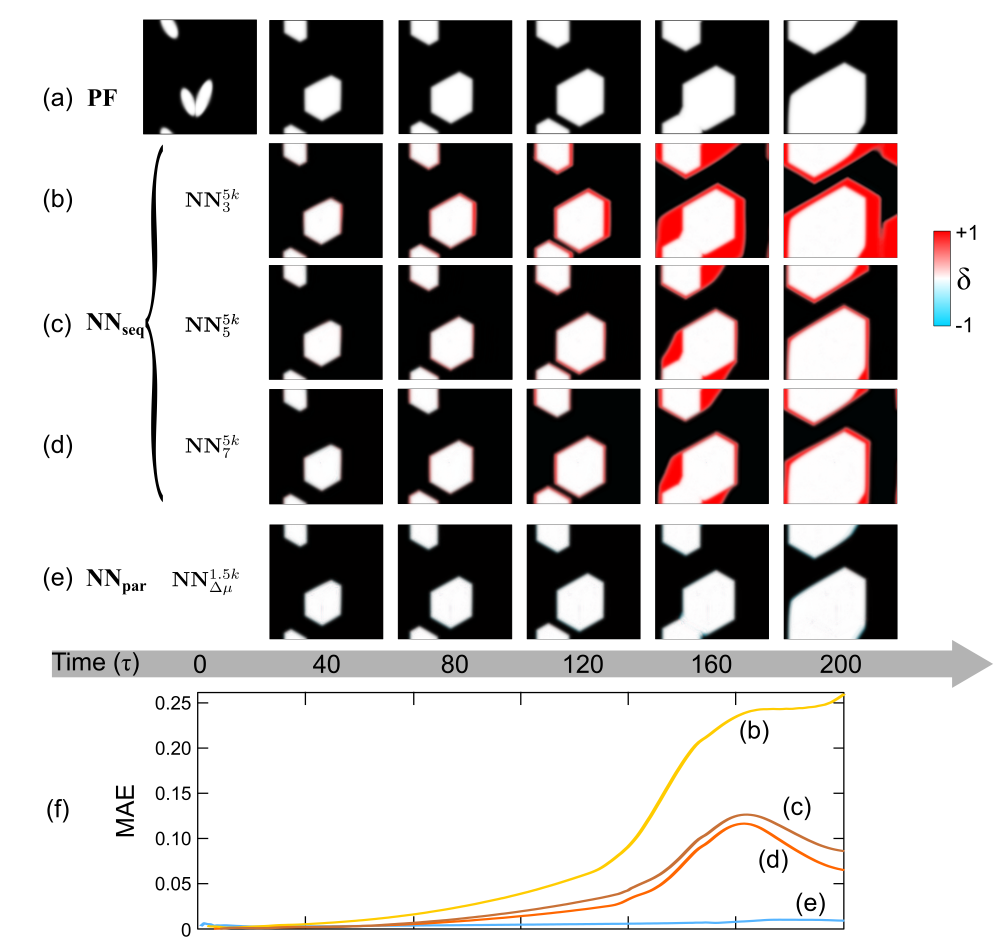}
    \caption{Evolution sequence for the same case of Fig.~5 of the main manuscript ($\Delta\mu=0.6)$ comparing the ground-truth PF solution (a) with the corresponding predictions from both \NNseq (b,c,d) and \NNpar (e) models. The pixel-by-pixel error $\delta$ is superimposed on the predicted images by colour bar. (f) Time evolution of the MAE between NN predictions and the true PF solution.}
    \label{fig::S4}
\end{figure}

\begin{figure}[ht!]
    \centering
    \includegraphics[width=\textwidth]{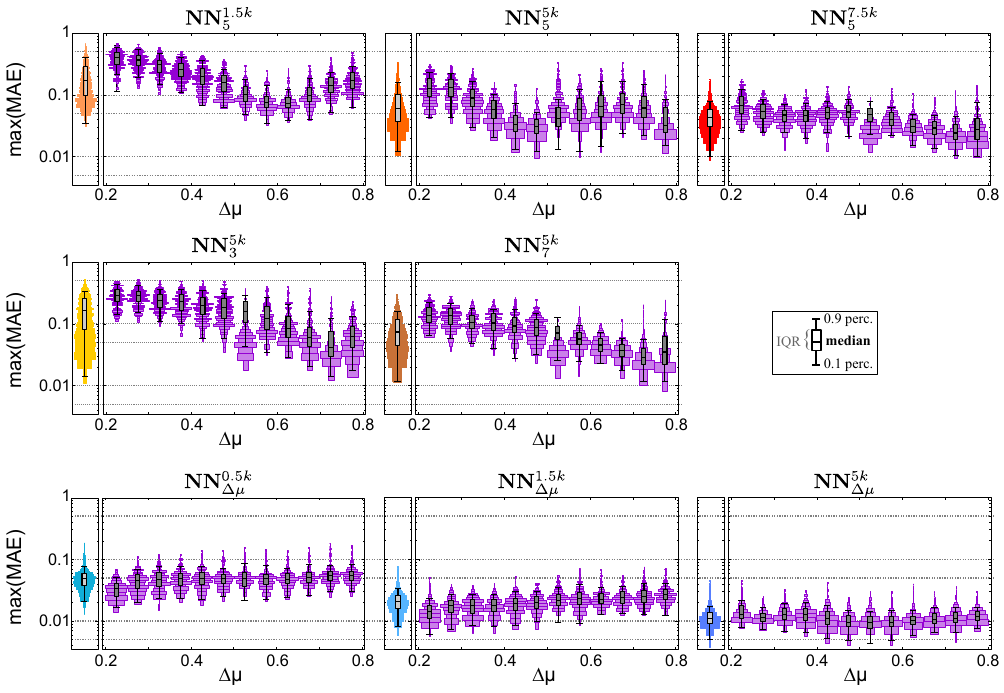}
    \caption{Analysis of the distribution of MAE maxima on the test set of $2500$ sequences used for Fig. 5 of the main manuscript, for all \NNseq and \NNpar trained models. For each of them, we report the overall distribution (first column) and those obtained by partitioning the test-set cases according to $\Delta\mu$ (binning width of $0.05$). Both violin plots and box-plots, reporting median, interquartile range (IQR) and the $[0.1,0.9]$ percentile range, are shown.}
    \label{fig::S5}
\end{figure}

\begin{figure}[ht!]
    \centering
    \includegraphics[width=\textwidth]{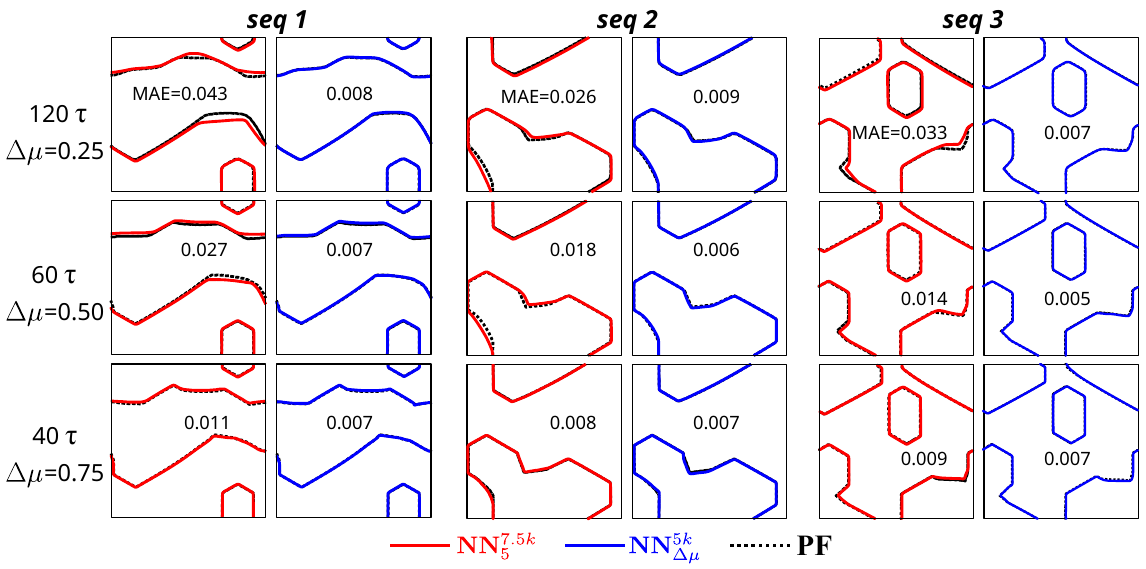}
    \caption{Comparison between true PF and NN-predicted profiles, for equivalent growth stage for simulations started from 3 different initial configurations ({\it seq 1} is the same as Fig.~7(b-d) of the main manuscript) and different $\Delta\mu$. Reported profiles correspond to the $\varphi=0.5$ contour lines. The MAE prediction error for each case is also indicated.}
    \label{fig::S6}
\end{figure}

\begin{figure}[ht!]
    \centering
    \includegraphics[width=0.9\textwidth]{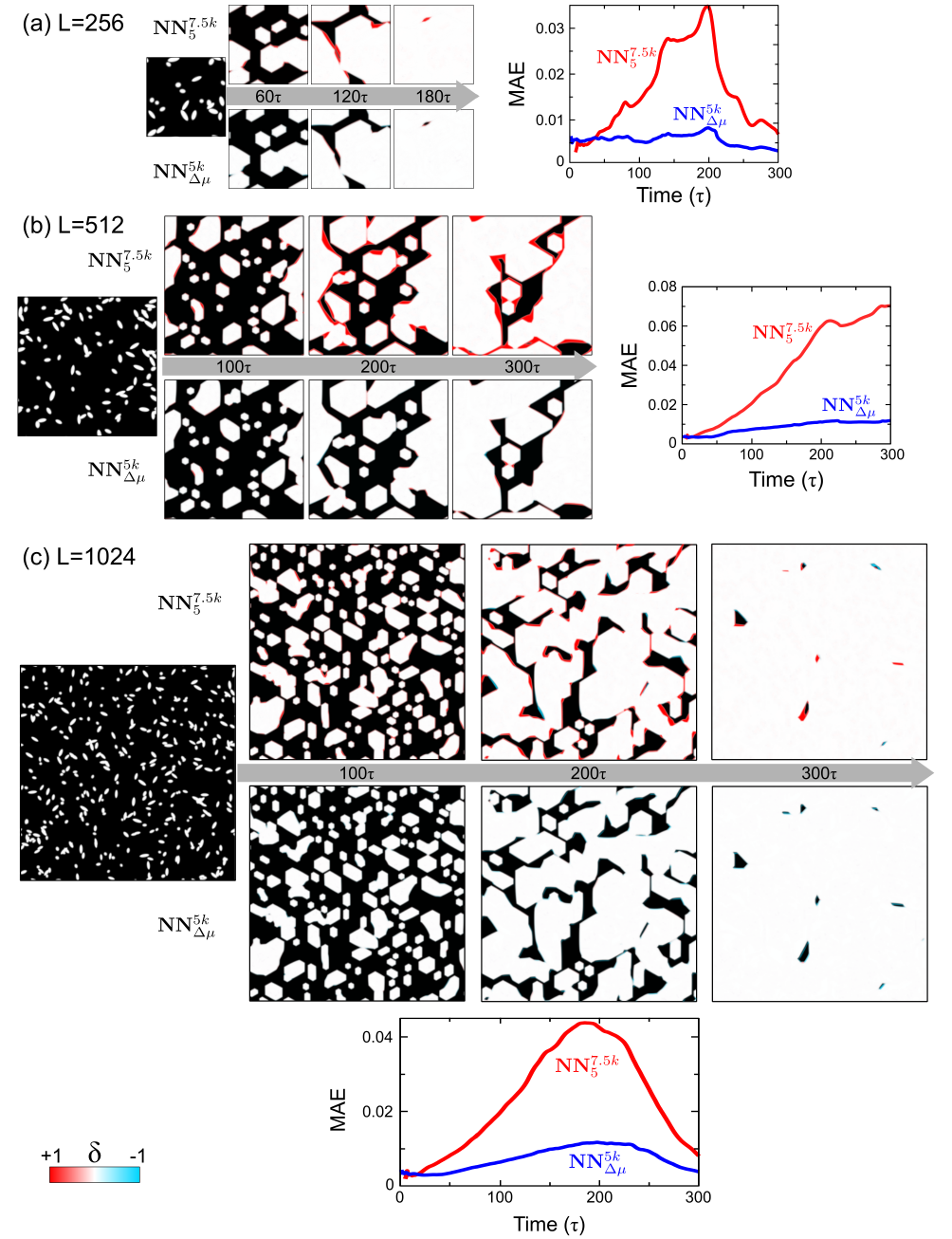}
    \caption{Evolution sequences for crystal growth under the same conditions of the training set but on different domain sizes: (a) $256 \times 256$ ($\Delta\mu\approx 0.58$), (b) $512 \times 512$ ($\Delta\mu\approx 0.23$) and (c) $1024 \times 1024$ ($\Delta\mu\approx 0.31$) as obtained by models \NNs{5}{7.5k} and \NNp{5k}. Pixel-by-pixel errors $\delta$ are superimposed on the evolution frames by color map. The corresponding evolutions of MAE over time are plotted.}
    \label{fig::S7}
\end{figure}

\begin{figure}[t!]
    \centering
    \includegraphics[width=\textwidth]{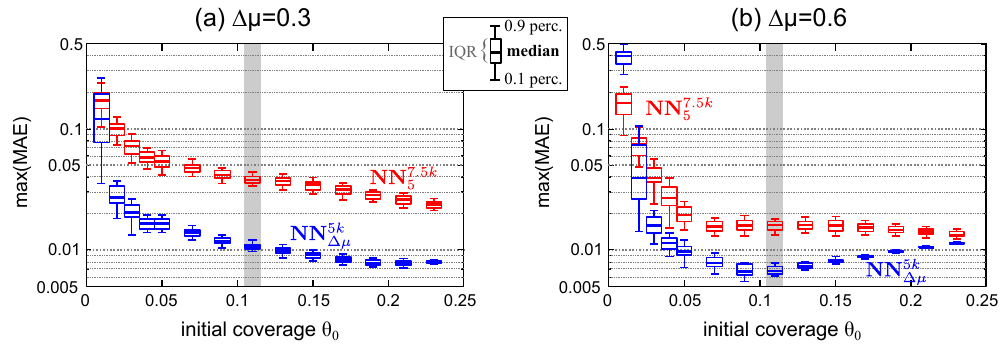}
    \caption{Dependence of the MAE maxima distribution as a function of the initial profile coverage $\theta_0$ for the same test set of $50$ evolution sequences used in Fig. 9(e) of main manuscript but with (a) $\Delta\mu=0.3$ and (b) $\Delta\mu=0.7$  Box-plots show the median, interquartile range (IQR) and $[0.1,0.9]$ percentile range. The gray shaded area corresponds to the training set coverage range.}
    \label{fig::S8}
\end{figure}

\end{document}